\documentclass[letterpaper, 10 pt, conference]{ieeeconf}  % Comment this line out
\IEEEoverridecommandlockouts                              % This command is only
\usepackage{graphics,color,epsfig,xspace,enumerate,multirow,subfigure}
\usepackage{graphicx}

\title{\LARGE \bf
Distributed Fair Scheduling Using Variable Transmission Lengths in Carrier-Sensing-based Wireless Networks
}

\author{Libin Jiang and Jean Walrand\\Dept. Electrical Engineering and Computer Sciences, University of California, Berkeley\\{\tt\small \{ljiang, wlr\}@eecs.berkeley.edu}.\thanks{* This work was supported in part by NSF grants NeTS-WN 0435478 and NeTS-FIND 0627161.}}

\begin{document}

\maketitle
\thispagestyle{empty}
\pagestyle{empty}

%%%%%%%%%%%%%%%%%%%%%%%%%%%%%%%%%%%%%%%%%%%%%%%%%%%%%%%%%%%%%%%%%%%%%%%%%%%%%%%%
\begin{abstract}

The fairness of IEEE 802.11 wireless networks (including Wireless LAN and Ad-hoc networks) is hard to predict and control because
of the randomness and complexity of the MAC contentions and dynamics. Moreover, asymmetric channel
conditions such as those caused by capture and channel errors
often lead to severe unfairness among stations.
In this paper we propose a novel distributed scheduling algorithm that we
call VLS, for ``{\em variable-length scheduling}'', that provides weighted fairness to
all stations despite the imperfections of the MAC layer and physical channels.
Distinct features of VLS include the use of variable transmission
lengths based on distributed observations, compatibility with 802.11's
contention window algorithm, opportunistic scheduling to achieve high throughput
in time-varying wireless environments, and flexibility and ease of implementation. Also, VLS makes the throughput of each station more smooth, which is appealing to real-time applications such as video and voice. 
Although the paper mostly assumes 802.11 protocol, the idea generally applies to wireless networks based on CSMA (Carrier Sensing Multiple Access).

\end{abstract}


%%%%%%%%%%%%%%%%%%%%%%%%%%%%%%%%%%%%%%%%%%%%%%%%%%%%%%%%%%%%%%%%%%%%%%%%%%%%%%%%
\section{INTRODUCTION}

In this paper, we propose a simple distributed scheduling algorithm
that provides weighted fairness in IEEE 802.11 \cite{80211} wireless
networks, despite the unpredictability of the 802.11 MAC layer and physical
channels.

In 802.11 wireless networks, MAC-layer contention, dynamics
and bandwidth allocation are hard to predict. For such networks, the
fixed-point model in \cite{Bianchi} gives a method for computing the
long-term throughput of the Binary Exponential Backoff (BEB) algorithm
\cite{80211}. However, the short-term dynamics and unfairness are quite unpredictable. In a certain period, a station may randomly
backoff more than others, and therefore have a smaller chance of winning the
channel, which in turn makes that station backoff even more.

Meanwhile, the BEB amplifies the unfairness caused by the impairments of the wireless channels. This aggravation is an unintentional side
effect of BEB that was designed to reduce collisions, not to guarantee fairness. The following two effects 
cause the unfairness:

\begin{itemize}

\item [(1)] {\em Capture:} Capture occurs when the signals from different
transmitters have very different strengths at a receiver \cite{Capture}.
For instance, a ratio of 2 in distances from the stations to the AP can lead to approximately
a ratio of 16 in received signal strengths. When more than one
station transmit packets to the AP at the same time, the AP may be able to capture and
correctly decode the packet from the closer station, while ignoring
the other packets. This effect increases
the aggregate throughput since the AP receives one packet even when multiple transmissions
overlap in time. However, capture may result in unfairness since the stations that are further away backoff
more with the BEB algorithm, and consequently obtain much less throughput than closer stations
\cite{sniff_capture}.

\item [(2)]  {\em Channel errors:} In addition to packet collisions, channel errors are another
important cause of packet loss. A more lossy channel to the AP drops more packets because of channel errors.
The transmitting station interprets all packet losses as collisions
and doubles its contention window. Accordingly, the BEB algorithm magnifies
the asymmetry of the lossy channels.
To alleviate this problem, reference \cite{Pang_loss_differentiation}
describes a way to differentiate the two kinds of packet losses (due
to collisions or channel errors). The algorithm proposed in
this paper provides a simpler solution.

\end{itemize}

With a more complicated MAC, IEEE 802.11e \cite{80211e} provides
Differentiated Service (DiffServ), by adopting different minimum Contention
Windows ($CW_{min}$) and inter-frame Spaces (IFS) for different service classes
such as voice, video and data. This protocol provides relative performance
differentiation among different classes: the classes with smaller $CW_{min}$ 
and IFS have a relative priority over others. To evaluate the performance of
802.11e, reference \cite{EDCF_sim} provides a simulation study; while
reference \cite{Saturated_80211e} uses an analytical model (a Markov
chain) to find the saturated throughput of 802.11e. However, the model
there is quite complicated, indicating that the {}``amount'' of
relative priority is hard to quantify and control. For instance, it is
not clear how much more bandwidth the protocol gives to video with a particular setting of $CW_{min}$ and IFS, nor
how to adjust the amount of priority by varying these parameters.

In this paper, we describe a simple, easy to implement, distributed fair scheduling algorithm that we call
VLS, for``variable-length scheduling,'' to cope with the above problems.
VLS provides exact weighted fairness despite the unpredictability of the 802.11
MAC layer and physical channels.

%%%%%%%%%%%%%%%%%%%%%%%%%%%%%%%%%%%%%%%%%%%%%%%%%%%%%%%%%%%%%%%%%%%%%%%%%%%%%%%%
\section{Variable-Length Scheduling (VLS): Bringing Order To Random Access}

In this section, we assume that there is only one collision domain. That is, each
station can sense the transmissions of other stations. (We consider
the case of multiple collision domains in section \ref{sec:VLS-in-Multiple-CD}.)
There are two versions of the scheduling algorithm: without and with an access point (AP). 
The latter is
an adaptation of the former that utilizes the  AP to simplify the algorithm.

\subsection{\label{sub:Distributed-algorithm}Distributed algorithm}

\begin{figure}
\noindent \begin{centering}
\includegraphics[width=3.3in]{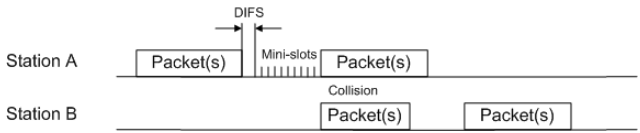}
\end{centering}
\caption{\label{fig:Virtual-slots}Virtual slots. (There are 3 virtual slots
in this figure.)}
\end{figure}

The algorithm is based on the concept of {}``virtual slot.'' By definition, a station
sees a virtual slot when it senses a collision, a burst of transmissions
(i.e., one DATA-ACK exchange, or a series of DATA-ACK exchanges separated
by SIFS), or when it transmits a burst of packets itself. Mini-slots
are not counted as virtual slots. In other words, a station counts a virtual slot whenever
if senses the channel as {}``idle'' for an interval equal to DIFS (DIFS$>$SIFS \cite{80211}) and is involved in a contention
process (i.e., when the station's backoff counter goes down until it transmits a
packet or senses other transmissions). In Fig. \ref{fig:Virtual-slots}, for example, there are
3 virtual slots. {}``Virtual slots'' are similar to {}``busy slots''
except that a burst of transmissions is counted as one {}``virtual
slot.''

The notion of {}``virtual slot'' is particularly useful because every station in
a single collision domain sees the same number of virtual slots, assuming
that the stations are always backlogged. (If not, the station starts the algorithm
only when it has a backlog and stops it when its backlog
is cleared.) Therefore, virtual slots can serve as a {}``clock'' for scheduling.
We design the distributed algorithm as follows.

\begin{itemize}
\item Each station $j$ is assigned a ``weight'' $W_j$ \cite{DFS}. (If there are multiple flows outgoing 
from station $j$, then let $W_j$ be the sum of the weights of all individual flows.) And each station $j$ keeps track of a value $m_j$ that is initially equal to 1.

\item If station $j$ gets an ACK after it transmits a packet, it keeps transmitting a burst of $m_j W_j$ packets separated by SIFS and then resets $m_j$ to the value 1. 

\item If it does not get an ACK after it transmits a packet, or if it does not get to transmit (i.e., it does not win a contention), station $j$ increments $m_j$ by one
whenever it sees a virtual slot.

\end{itemize}

\begin{figure}
\noindent \begin{centering}
\includegraphics[width=3.3in]{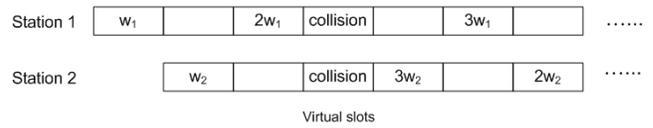}
\end{centering}
\caption{\label{fig:VLS_2sta}Variable-length Scheduling. ($w_1$, $w_2$ are the weights. Each block represents a variable-length virtual slot. The number in a block indicates the transmission length.)}
\end{figure}

For example, Fig. \ref{fig:VLS_2sta} shows the process with 2 stations with weights $w_1$ and $w_2$ respectively. In the figure, each block represents a virtual slot (aligned across different stations). The number in each block indicates the transmission length in that virtual slot. (Note that the size of a block here does not reflect the actual length of the virtual slot.) Assume station 1 starts transmission with the 1st virtual slot, while station 2 starts transmission with the 2nd.

In a time period when station $j$ is backlogged, we can see that
the total number of packets it has transmitted is equal to the number
of virtual slots it has seen so far times the weight $W_{j}$. (So station $j$ {\em virtually} transmits $W_{j}$ packets per slot. This is why we use the name ``virtual slot''.)
Since all stations see the same virtual slots, the bandwidth allocation
is weighted fair. That is $T_{j}/W_{j}$ is the same for all the stations where
$T_{j}$ is the average rate of packets that station $j$ transmits.

Note that VLS guarantees the weighted fairness, no matter what happens
in the MAC and physical layers. Thus, VLS automatically adjusts for
the randomness of the MAC protocol and the asymmetry of the physical channels.

\subsection{Algorithm with an AP}

In a wireless LAN with an AP, the above algorithm can be adapted so
that the client stations need not count the virtual slots. In this variation, 
the AP counts the virtual slots for the stations and piggybacks
that count in the MAC-layer ACKs to the stations. The algorithm
works as follows.

\begin{itemize}
\item The AP keeps counting the virtual slots. It increments the count ($v++$) for each virtual slot.
\item A station can start and stop the algorithm at any time. When station
$j$ starts, it contends for the channel and sends the first $W_{j}$
packets to the AP. In one ACK, the AP piggybacks the current value
of $v$, denoted as $v_{i}^{j}$. The next time station $j$ wins
the channel, it sends $W_{j}$ packets first. The AP, again, piggybacks
the current value of $v$, denoted as $v_{i+1}^{j}$. Then, the station
sends $W_{j}(v_{i+1}^{j}-v_{i}^{j}-1)$ more packets in the burst.
Since $v_{i+1}^{j}-v_{i}^{j}-1\ge0$, station $j$ sends at least
$W_{j}$ packets per burst.
\end{itemize}

\subsection{Considerations on Burst Length}

Suppose that, at each virtual slot, every station $i$ has a probability
$p_{i}$ of winning the channel. Then on average, station $i$ accumulates
$W_{i}/p_{i}$ units of {}``credit'' before it wins the channel
(since the average number of virtual slots it waits for is $1/p_{i}$).
It can then spent the credits by transmitting (on average) $W_{i}/p_{i}$
packets. When the number of active stations $N$ increases in a wireless
networks, $p_{i}$ decreases (approximately $\propto1/N$). As
a result, the average burst length increases, thus causing more
delays for all the stations.

To avoid this effect, we define a system-wide parameter $c>0$, called ``{\em speed of the clock}'',  and we modify the protocol as follows:

\begin{itemize}
\item Instead of transmitting $m_{j}W_{j}$ packets in a burst as in subsection
\ref{sub:Distributed-algorithm}, station $j$ transmits $c\cdot m_{j}W_{j}$
packets. If $c\cdot m_{j}W_{j}$ is not an integer, then it transmits
$\left\lfloor c\cdot m_{j}W_{j}\right\rfloor $ packets, and saves
the extra credits $c\cdot m_{j}W_{j}-\left\lfloor c\cdot m_{j}W_{j}\right\rfloor $
for the next time. Essentially, $c$ controls the speed of clock in
the whole network. (Therefore, the delay is proportional to $c$.)
\item The stations can adjust the value of parameter $c$ in several different ways:

\begin{itemize}
\item If station $j$ knows the number of backlogged stations $N$, then
it can choose $c_{j}=1/N$ and broadcast $c_{j}$ to the network.
The other stations will then follow the parameter. 
\item Common TCP flows are usually not sensitive to the burst length and
delay. But if a station (say, station $j$) has delay-sensitive flows
and some other stations' burst lengths are causing too much delay
to it, it computes a new value of $c_{j}$ and broadcasts it to the
network. (Assume the current average delay for station $j$ is $d_{j}$,
and its targeted delay is $D_{j}$, then set $c_{j}=c_{0}\cdot D_{j}/d_{j}$,
where $c_{0}$ is the current parameter of the system.) 
\item If the network has an AP, the AP can act as a controller to adjust
$c$.
\item If there are more than one stations broadcasting $c_{j}$, each station
follow the lowest $c_{j}$ it has received (i.e., $c=\min_{j}c_{j}$).
\end{itemize}
\item Further details about the implementation of broadcasting: 

\begin{itemize}
\item Station $j$ embeds $c_{j}$ in a packet (or piggybacked in a usual
data packet), along with its ID/address. 
\item To increase reliability, this packet can be repeated multiple times.
Also, although stations in a single collision domain may {}``carrier-sense''
each other, they may not be able to {}``decode'' the packets of
each other. Therefore, the packets containing $c_{j}$ are transmitted
with a higher power, or a lower data rate, than usual packets. 
\item If a station has broadcast a $c_{j}$ before and wants to update it,
it simply broadcasts the new $c_{j}$. Since other stations know the
ID of the sender, they update the old parameter of the same sender,
and follow the lowest $c_{j}$ in their records.
\end{itemize}
\end{itemize}
In addition, we can impose a limit on the burst length, $B_{j}$, of each
station $j$. In this case, station $j$ can transmit up to $min(B_{j},c\cdot m_{j}W_{j})$
packets in a burst (and the remaining credit is left for future transmissions). This mechanism smooths out the randomness of the burst lengths,
which may otherwise be (randomly) long or short, even if the
network has a proper value of $c$. But in this situation, $c$ needs to be
small enough to avoid the instability of credits (i.e., the remaining credits 
should not go to infinity). In particular, a
simple inequality needs to be satisfied. We discuss this issue
in Section \ref{sec:Stability}.

\subsection{\label{sub:Generalized-Fair-Scheduling}Generalized Fair Scheduling}

As mentioned before, virtual slots act as a clock for scheduling.
Using this synchronization mechanism, VLS has the flexibility to achieve many forms of
fairness. The scheduler above uses the number of packets as a fairness metric.
VLS can also provide weighted fairness in terms of the number of bits
or the {}``air-time'' occupied by different stations. If different
stations use very different data rates (e.g., 1Mbps vs. 11Mbps) in
a shared-medium wireless network, providing fairness in terms of bits
leads to very low efficiency (throughput) of the whole network \cite{PF_Jiang}.
In this case, \cite{PF_Jiang} shows that allocating equal air-time
to different stations strikes a good balance between fairness and
efficiency, and is actually equivalent to achieving proportional fairness
\cite{kelly97charging}.

%%%%%%%%%%%%%%%%%%%%%%%%%%%%%%%%%%%%%%%%%%%%%%%%%%%%%%%%%%%%%%%%%%%%%%%%%%%%%%%%
\section{Performance Evaluation}

\subsection{Short-term fairness}

\begin{figure}
%\subfigure[w/o VLS]{\includegraphics[scale=0.5]{unfair}}
\subfigure[w/0 VLS]{\epsfig{figure=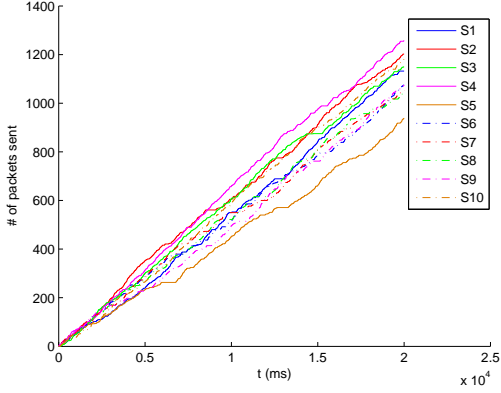,scale=0.5}}
%\subfigure[VLS]{\includegraphics[scale=0.5]{fair.eps}}
\subfigure[VLS]{\epsfig{figure=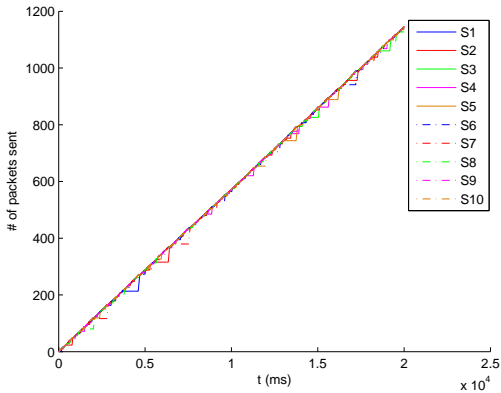,scale=0.5}}

\caption{\label{fig:short-term}Improvement of short-term and long-term fairness
(10 stations, $W_{j}=1,\forall j$) (Throughout the paper, ``S1'' means ``Station 1'', etc.)}
\end{figure}

In a wireless LAN, or an ad-hoc network with a single collision domain,
the long-term saturated throughput should the same for all stations, by symmetry \cite{Bianchi}.
However, the collisions and the dynamics of contention windows (with
BEB) are quite unpredictable, leading to fluctuations of short-term
throughput (Fig. \ref{fig:short-term}(a)). Also, the volumes of data
that different stations send drift away from each other (Fig. \ref{fig:short-term}(a)).
This means that, although the average throughput of the different stations
are equal in the long term, the average throughputs may differ over a
considerable time window (20 sec in the figure). With VLS, short-term
and long-term fairness have been clearly improved (Fig. \ref{fig:short-term}(b)).

\subsection{Weighted Fairness}

\begin{figure}
\subfigure[w/o VLS (The vector of $CW_{min}$'s of the 10 stations is (128 64 128 64 42 32 128 64 26 42))]{\includegraphics[scale=0.5]{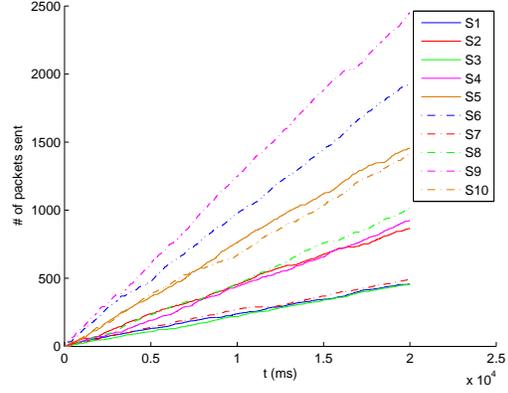}}
\subfigure[VLS (The vector of weights of the 10 stations is (1 2 1 2 3 4 1 2 5 3))]{\includegraphics[scale=0.5]{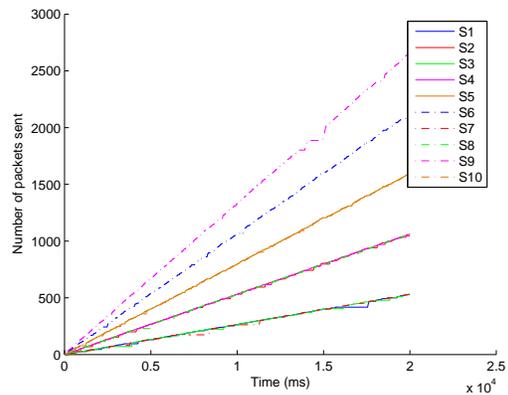}}

\caption{\label{fig:Weighted-Fairness}Weighted Fairness}
\end{figure}

Without VLS, weighted fairness may be implemented by using different
$CW_{min}$'s. Approximately, the throughput of an individual station is inversely
proportional to its minimum CW, assuming that each station has the same average packet
size and use the same IFS (Fig. \ref{fig:Weighted-Fairness}(a)). However, the approximation
is not accurate (especially when $CW_{min}$'s are small), and one can expect
it to be vulnerable to physical layer factors such as capture effect
and channel errors. With VLS, the weighted fairness is exact, and easy
to adjust and control (Fig. \ref{fig:Weighted-Fairness}(b)).

\subsection{Solving Unfairness Problems}

Besides improving short-term fairness and providing weighted fairness,
our algorithm can readily solve many other unfairness problems in
wireless networks.

\subsubsection{Unfairness due to Capture Effect}

\begin{figure}
\subfigure[w/o VLS]{\includegraphics[scale=0.5]{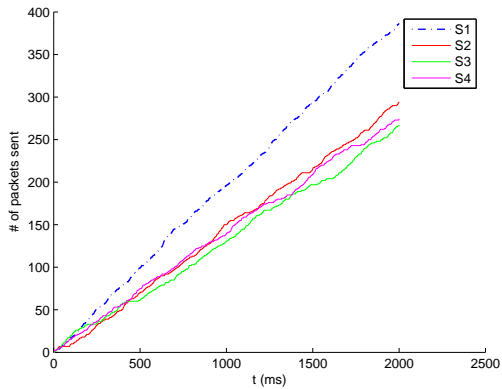}}
\subfigure[VLS]{\includegraphics[scale=0.5]{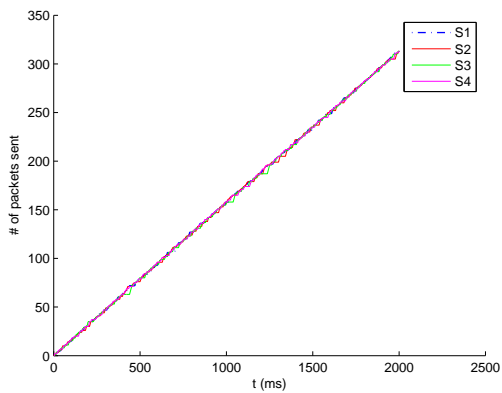}}

\caption{\label{fig:capture}Overcoming unfairness due to capture effect (S1 is the strong station. When S1's packets collide with other stations' packets, S1's packet is captured)}

\end{figure}

In a WLAN, different stations may have very different distances from
the AP. Also, the channel qualities between stations and the AP may
differ greatly even if the distances are similar, for instance because of multipath
or obstructions. The above effects result in different signal strengths
from different stations as received by the AP. When more than one
stations transmit packets to the AP at the same time , so that these
transmissions collide, the AP may still be able to
capture and correctly decode the packet with the strongest signal,
and send back an ACK. This feature is helpful in terms of the aggregate
throughput since one packet is received even in the event of a collision, but may
exacerbate the unfairness. The weaker stations tend to backoff more
with the BEB algorithm, and therefore obtain much less throughput than stronger ones
(see Fig. \ref{fig:capture}(a)).

If our VLS algorithm is enabled, it can overcome the unfairness problem,
as well as \emph{retaining the throughput benefit} provided by capture. Since both strong and weak stations
share the same view of the virtual slots, they share the bandwidth
in a fair way (see Fig. \ref{fig:capture}(b)).

\subsubsection{Unfairness due to Channel Errors}

\begin{figure}
\subfigure[w/o VLS, only S1 suffers a loss rate $p$=0.15 due to channel errors]{\includegraphics[scale=0.5]{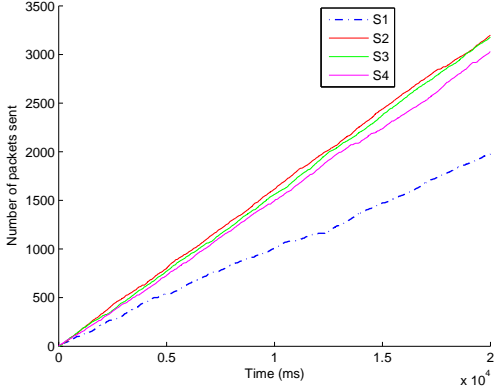}}
\subfigure[With VLS, $p$=0.15 for S1]{\includegraphics[scale=0.5]{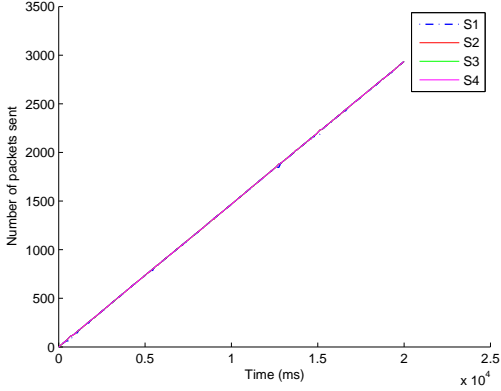}}
\subfigure[With VLS, $p$=0 for all stations]{\includegraphics[scale=0.5]{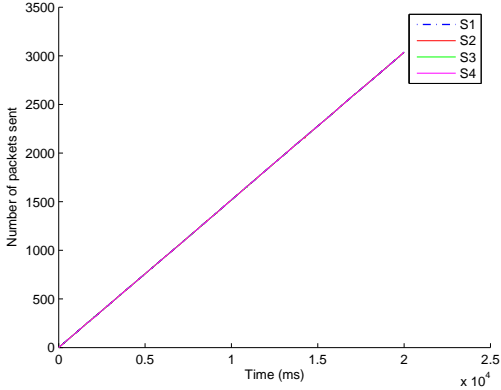}}

\caption{\label{fig:channel error}Overcoming unfairness due to channel errors (Note that in (b) fairness is achieved, also, the throughput is close
to (c) w/o channel errors.)}

\end{figure}

In wireless networks, in addition to packet collisions, channel errors are another
important cause of packet losses. If a station has a lossy
channel to the AP, its packets are dropped with higher probability
because of channel errors. Similarly to the capture effect, these losses also result in
an asymmetry among different stations, and therefore in unfairness, aggravated
by the BEB algorithm. This effect is shown in Fig. \ref{fig:channel error}(a) where
only station 1 suffers from a loss probability $p=0.15$ caused by
channel errors.

\begin{figure}
\noindent \begin{centering}
\includegraphics[width=1.5in]{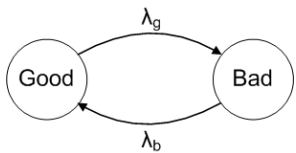}
\par\end{centering}

\caption{\label{fig:Markov-Channel-Model}Markov Channel Model}
\end{figure}

A simple model of a time-varying wireless channel is a Markov chain with two
states: {}``good'' and {}``bad''. For simplicity, we assume that
in the {}``good'' state all packets can be received and that in the {}``bad''
state all packets are dropped. Each state has an exponentially-distributed
duration before it transits to the other state. Fig. \ref{fig:Markov-Channel-Model}
shows the state transition diagram. In the scenario simulated, only
station 1 has such a noisy channel, with $\lambda_{g}=20/sec$, $\lambda_{b}=113/sec$
(therefore the average loss probability is $p=0.15$); and other stations
have perfect channels ($p=0$). It turns out that our algorithm not
only maintains fairness, but also utilize the channel {}``\emph{opportunistically}''
to get a high throughput: When the weak station wins a contention,
but meets the {}``bad'' state, it transmits a packet, without receiving
the ACK, and relinquishes the channel immediately. When it wins a
contention and meets the {}``good'' state, it can transmit a burst
of packets. Since the channel is likely to stay in the {}``good''
state for some time, the station has the opportunity to compensate
for its past losses. If at some point the state goes back to {}``bad'',
the station stops immediately and waits for the next opportunity (see
Fig. \ref{fig:channel error}(b)). In other words, the weak station transmits
more when the channel is good, and less when the channel is bad, thus utilizing
the noisy channel more efficiently. As a result, the total throughput
in Fig. \ref{fig:channel error}(b) is only slightly less than that
in Fig. \ref{fig:channel error}(c), where the channels are perfect
for all stations.

It is evident that if the average duration of the {}``good'' state
is too short, the weak station may still receive unfair throughput.
We can derive the condition under which fairness is guaranteed (see
Section \ref{sec:Stability}). Nevertheless, VLS always improves the fairness.

When the channel of one station is significantly worse than that of the others,
providing throughput-fairness to different stations may drag down
the total throughput of the network. (This is a tradeoff between fairness
and efficiency.) In that case, providing time-fairness would be more
suitable. VLS has the flexibility to provide time-fairness, as mentioned
in subsection \ref{sub:Generalized-Fair-Scheduling}.

In the above, we have assumed that the stations do not know the state
of the channel before transmitting. If the channel state information
is known in advance, the station can avoid transmitting in the {}``bad''
state. In that case, the throughput performance is further improved.

%%%%%%%%%%%%%%%%%%%%%%%%%%%%%%%%%%%%%%%%%%%%%%%%%%%%%%%%%%%%%%%%%%%%%%%%%%%%%%%%

\section{\label{sec:Stability}Stability of the Credits}

In VLS, each active station keeps accumulating and spending credit, where the accumulated credit is proportional to a station's weight, the number of virtual slots it has observed, and the speed of clock $c$. In this section, we consider whether the credits are stable, that is,
whether the credits of some stations keep increasing and go to infinity.
If that happens, the bandwidth allocation may not be fair, since the
extra credits of some stations are not spent.

First, if there is no limit on the length of transmission at each
burst, the credits must be stable, given that each station has non-zero
probability of winning the channel. Suppose at each virtual slot,
a station $i$ has a probability of $p_{i}$ of winning the channel,
then on average, it has accumulated $W_{i}\cdot c/p_{i}<\infty$ units
of credit. It then spent all of them. In fact, the probability that
the credits reach $G$ (before it is spent) is approximately $(1-p_{i})^{G/(W_{i}\cdot c)}$.
Therefore $Pr\{ G=\infty\}=0$.

But if there is a pre-defined limit of the burst length $B_{i}$ for
station $i$, then instability is possible. In particular, if $B_{i}\le W_{i}\cdot c/p_{i}$,
the credits of station $i$ go to infinity. This can be readily proved
by law of large numbers. Also, this inequality tells us how to avoid
instability: ensuring \begin{equation}
B_{i}>W_{i}\cdot c/p_{i},\forall i\label{eq:stability}\end{equation}

For implementation, each station monitors its $p_{i}$ (average
over a period of time). If the inequality is violated, it computes
a proper value of $c$ and broadcasts it to the network. Then every station
uses the new value of $c$. Another simple implementation
is to monitor the credits. If the credits of station $i$ keep increasing,
it knows a smaller $c$ is needed to stabilize its credit.

The above analysis can be extended to the case of capture effect or
channel errors.

\begin{enumerate}
\item Capture effect. Capture effect will affect the values of the $p_{i}$'s. The weaker
stations have smaller $p_{i}$'s, which, in turn, may entail adjustment
of $c$.
\item Channel errors. Here, we define $p_{i}$ as station $i$'s probability
of winning the contention AND meet the {}``good'' state of channel
$i$, in a given virtual slot. Accordingly, channel errors clearly affect the $p_{i}$'s.
The stations with noisy channels have smaller $p_{i}$'s (due to both
channel errors and BEB). Also, the time-variation of channel quality
imposes another limit on the burst length. Denote the length of the
{}``good'' period of channel $i$ as $T_{i}$, which is a random
variable. Then the following inequality is required:\begin{equation}
{\bf E}[min(B_{i},T_{i})]>W_{i}\cdot c/p_{i},\forall i\label{eq:stability with channel errors}\end{equation}

\end{enumerate}
To analyze the values of $p_{i}$'s, one can adapt Bianchi's fixed
point model \cite{Bianchi}. In practice, the stations do not need
to compute $p_{i}$'s. They only need to adjust $c$ according to
their extra credits.

%%%%%%%%%%%%%%%%%%%%%%%%%%%%%%%%%%%%%%%%%%%%%%%%%%%%%%%%%%%%%%%%%%%%%%%%%%%%%%%%

\section{\label{sec:VLS-in-Multiple-CD}VLS in Multiple Collision domain}

If a wireless network has multiple collision domains, a station may
not be able to hear all the other stations' transmissions. Therefore,
different stations may have different views of the virtual slots, 
which makes virtual-slot-based scheduling more difficult. (Two
stations have the same view of virtual slots only if they can hear
the same set of stations.) In IEEE 802.11, this may cause severe unfairness problem 
(as will be shown later). So, in this case, we devise a variable-length
scheduling algorithm based on the realized throughputs, instead of
the number of virtual slots. The basic idea is similar to \cite{fair_by_observation}\cite{Gupta_fairness}.
(There, the minimum Contention Windows, $CW_{min}$'s, are dynamically adjusted.)

Say we have a set of $J$ stations, with respective weights $W_{j},j=1,2,\dots,J$.
The weights can be pre-determined by optimizing some global objective
of the network. For example, they can be the solution of the utility
optimization problem \cite{kelly97charging}\cite{tutorial_wireless}
\begin{eqnarray*}
 & max & \sum_{j}U_{j}(W_{j})\\
 & st & \sum_{k\in{\cal C}_{m}}W_{k}\le1,\forall{}_{m},m=1,2,\dots,M\end{eqnarray*}
where $U_{j}$ is the concave {}``utility function'' of station
$j$, and each set ${\cal C}_{m}$ is a {}``clique'' (in a clique,
only one station can transmit at a time). Note that with the constraints, we
have assumed that the contention graph is a ``perfect graph'' \cite{ad_hoc_constraints} (otherwise, another set of constraints based on ``independent set'' should be used \cite{ad_hoc_constraints}).
We have also omitted some details such as packet collisions. This optimization
problem can be solved in a distributed manner, similar to wired network
\cite{kelly97charging}. A {}``clique'' here is analogous to a {}``link''
in wired network, therefore to solve the problem, some communication
among stations in the same clique is needed.

In this VLS algorithm, each station $j$ monitors the aggregate throughput
of its neighbors ${\cal N}_{j}$ (i.e, $\sum_{k\in{\cal N}_{j}}S_{k}$),
as well as its own throughput $S_{j}$. (This can be done in several
ways: (a) station $j$ can overhear the packets sent by its neighbors,
if possible; (b) otherwise, stations can explicitly exchange information
about their throughput with their neighbors periodically.) Then, it
adjusts the burst length as follows.\[
b_{j}(t+1)=b_{j}(t)-\alpha(t)b_{j}(t)(\frac{S_{j}}{\sum_{k\in{\cal N}_{j}}S_{k}}-\frac{W_{j}}{\sum_{k\in{\cal N}_{j}}W_{k}})\]
where $b_{j}(t)$ denotes the burst length of station $j$ at time
$t$, and $\alpha(t)$ is the step size. Clearly, $b(t)$ converges
when the actual ratio of throughputs is equal to the target ratio
of weights.

In the following simulation, we compare the throughput allocation
with fixed-length scheduling and VLS. In VLS, we use a discrete version
of the above algorithm: each node adjusts its transmission length
every 4ms, and the throughput $S_{j}$ is an average over the last
40ms.

\begin{figure}
\noindent \begin{centering}
\subfigure[A network with 2 collision domains]{\includegraphics[width=2in]{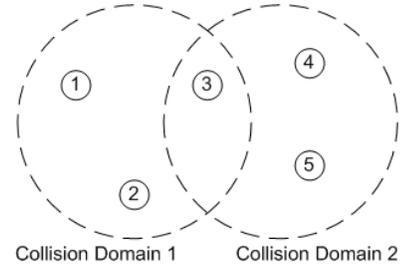}}
\par\end{centering}

\subfigure[Without VLS]{\includegraphics[scale=0.5]{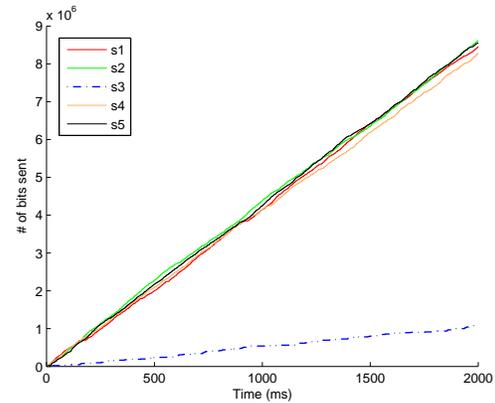}}
\subfigure[VLS, with equal weights for all stations]{\includegraphics[scale=0.5]{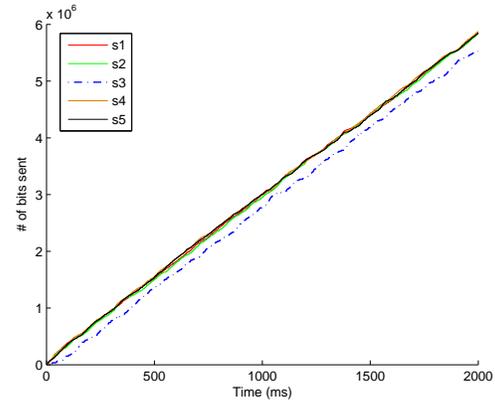}}

\caption{\label{fig:VLS-in-Multiple}VLS in Multiple Collision Domains}
\end{figure}

In the network simulated, stations 1, 2, 3 belong to Collision Domain
1, while stations 3, 4, 5 belong to Collision Domain 2, as shown in
Fig. \ref{fig:VLS-in-Multiple}(a). Note that station 3 faces contentions
from both domains. The destinations of the flows from node 1, 2, 
3, 4, 5 are assumed to be node 2, 1, 2, 5, 4, respectively. 
The data rate is 11Mbps, and the initial transmission
length is 1ms for all stations.

Without VLS, station 3's throughput is very low compared to others
(Fig. \ref{fig:VLS-in-Multiple}(b)). The reason is that at most of
the time, station 3 senses the medium as busy, due to the transmissions
in both collision domains. Therefore it does not have much chance
to transmit its packets. Then, we use VLS, and require the weights
of all stations to be 1/3. As shown in Fig. \ref{fig:VLS-in-Multiple}(c),
after a short period of time in the beginning (for convergence), the
target weights are achieved. On average, station 3' transmission length
is 1.54ms, while others' are about 0.5ms after convergence.

%%%%%%%%%%%%%%%%%%%%%%%%%%%%%%%%%%%%%%%%%%%%%%%%%%%%%%%%%%%%%%%%%%%%%%%%%%%%%%%%%%%%%%%%%%%%%%%
\section{Conclusions}

In this paper, we have proposed a distributed algorithm, VLS, for
fair scheduling in 802.11 wireless networks. We have shown that, by
varying the transmission lengths of different stations, it is feasible
to provide fairness without careful modeling of MAC-layer contention,
physical channel variations, and the effect of multiple collision
domains. Many existing fairness problems in 802.11 networks are therefore
overcome, including short-term unfairness, unfairness introduced by
physical channel's asymmetry, such as capture effect and channel errors. 
It makes the throughput of each station more smooth, which is appealing to real-time service such as video and voice. It can easily provide weighted fairness for 
different services. For networks with multiple collision domains, VLS gives a way to avoid the starvation of those stations that are in several collision domains.

Since VLS can avoid excessively long transmissions by tuning the parameter
$c$, we should be able to achieve certain objectives on packet delay.
However, this problem has not been studied thoroughly in this paper,
and is a subject for further study.

Since VLS can be used in 802.11e networks, the advantages of both 
protocols can be achieved: VLS can provide weighted fairness (a prioritization) in terms of throughput, while 802.11e can provide prioritization in terms of delay, by using different $CW_{min}$'s and IFS's (Inter-Frame Spaces). Therefore, throughputs and delays can be controlled separately by two protocols, instead of being coupled in a complicated way as in 802.11e \cite{Saturated_80211e}. More analytical/experimental study of this issue is interesting for future research.
%%%%%%%%%%%%%%%%%%%%%%%%%%%%%%%%%%%%%%%%%%%%%%%%%%%%%%%%%%%%%%%%%%%%%%%%%%%%%%%%
%\section{ACKNOWLEDGMENTS}

%%%%%%%%%%%%%%%%%%%%%%%%%%%%%%%%%%%%%%%%%%%%%%%%%%%%%%%%%%%%%%%%%%%%%%%%%%%%%%%%

\bibliographystyle{plain}
%\bibliography{wireless}

\begin{thebibliography}{10}

\bibitem{80211}
Information technology- telecommunications and information exchange between
  systems- local and metropolitan area networks- specific requirements- part
  11: Wireless lan medium access control (mac) and physical layer (phy)
  specifications.
\newblock {\em ANSI/IEEE Std 802.11, 1999 Edition (R2003)}, pages i--513, 2003.

\bibitem{80211e}
Unapproved draft amendment standard for information technology--
  telecommunications and information exchange between systems--lan/man specific
  requirements-- part 11 wireless medium access control (mac) and physical
  layer (phy) specifications: Medium access control (mac) quality of service
  (qos) enhancements (replaced by approved draft 802.11e/d13.0).
\newblock {\em IEEE Std P802.11e/D13.0}, 2005.

\bibitem{fair_by_observation}
B.~Bensaou, Wang Yu, and Ko~Chi~Chung.
\newblock Fair medium access in 802.11 based wireless ad-hoc networks.
\newblock In {\em Mobile and Ad Hoc Networking and Computing, 2000. MobiHOC.
  2000 First Annual Workshop on}, pages 99--106, 2000.

\bibitem{Bianchi}
G.~Bianchi.
\newblock Performance analysis of the ieee 802.11 distributed coordination
  function.
\newblock {\em Selected Areas in Communications, IEEE Journal on},
  18(3):535--547, 2000.
\newblock 0733-8716.

\bibitem{Gupta_fairness}
Rajarshi Gupta and Jean Walrand.
\newblock {\em Achieving Fairness in a Distributed Ad-Hoc MAC}.
\newblock "Advances in Control, Communication Networks, and Transportation
  Systems," E.H. Abed (Ed.), Systems and Control: Foundations and Applications
  Series, Springer-Birkhauser, Boston. July 2005.

\bibitem{kelly97charging}
F.~Kelly.
\newblock Charging and rate control for elastic traffic.
\newblock {\em European Transactions on Telecommunications, 8:33--37}, January
  1997.

\bibitem{sniff_capture}
A.~Kochut, A.~Vasan, A.~U. Shankar, and A.~Agrawala.
\newblock Sniffing out the correct physical layer capture model in 802.11b.
\newblock In {\em Network Protocols, 2004. ICNP 2004. Proceedings of the 12th
  IEEE International Conference on}, pages 252--261, 2004.

\bibitem{Capture}
C.~T. Lau and C.~Leung.
\newblock Capture models for mobile packet radio networks.
\newblock {\em Communications, IEEE Transactions on}, 40(5):917--925, 1992.
\newblock 0090-6778.

\bibitem{PF_Jiang}
Jiang Li~Bin and Liew Soung~Chang.
\newblock Proportional fairness in wireless lans and ad hoc networks.
\newblock In {\em Wireless Communications and Networking Conference, 2005
  IEEE}, volume~3, pages 1551--1556 Vol. 3, 2005.

\bibitem{tutorial_wireless}
X.~Lin, N.B. Shroff, and R.~Srikant.
\newblock A tutorial on cross-layer optimization in wireless networks.
\newblock {\em Selected Areas in Communications, IEEE Journal on},
  24(8):1452--1463, 2006.

\bibitem{Pang_loss_differentiation}
Liew~S.C Qixiang~Pang, Leung~V.C.M.
\newblock Improvement of {WLAN} contention resolution by loss differentiation.
\newblock {\em IEEE Transactions on Wireless Communications}, 5:3605 -- 3615,
  December 2006.

\bibitem{ad_hoc_constraints}
John~Musacchio Rajarshi~Gupta and Jean Walrand.
\newblock Sufficient rate constraints for {QoS} flows in ad-hoc networks.
\newblock {\em AD HOC NETWORKS}, 5(4):429--443, May 2007.

\bibitem{Saturated_80211e}
J.~W. Robinson and T.~S. Randhawa.
\newblock Saturation throughput analysis of ieee 802.11e enhanced distributed
  coordination function.
\newblock {\em Selected Areas in Communications, IEEE Journal on},
  22(5):917--928, 2004.
\newblock 0733-8716.

\bibitem{EDCF_sim}
Choi Sunghyun, J.~del Prado, N.~Sai~Shankar, and S.~Mangold.
\newblock "{IEEE} 802.11 e contention-based channel access ({EDCF}) performance
  evaluation".
\newblock In {\em Communications, 2003. ICC '03. IEEE International Conference
  on}, volume~2, pages 1151--1156 vol.2, 2003.

\bibitem{DFS}
N.~Vaidya, A.~Dugar, S.~Gupta, and P.~Bahl.
\newblock Distributed fair scheduling in a wireless {LAN}.
\newblock {\em Mobile Computing, IEEE Transactions on}, 4(6):616--629, 2005.

\end{thebibliography}

\end{document}